\documentclass[twocolumn,preprintnumbers,amsmath,amssymb,prb,aps,superscriptaddress]{revtex4-1}
\usepackage{txfonts}
\usepackage{pifont}
\usepackage{amssymb}
\usepackage{dcolumn}
\usepackage{amsmath}
\usepackage[dvips]{epsfig}
\usepackage{array}
\usepackage{color}
\usepackage{ulem}
\usepackage{caption}
\usepackage{graphicx}
\usepackage{float}
\usepackage{subcaption}
\usepackage{titlesec}
\usepackage[colorlinks=true,linkcolor=blue,citecolor=blue,urlcolor=blue,]{hyperref}
\newcolumntype{C}[1]{>{\centering}p{#1}}
\makeatletter
\def\btt#1{\texttt{\@backslashchar#1}}%
\DeclareRobustCommand\bblash{\btt{\@backslashchar}}%
\makeatother

\begin{document}
\bibliographystyle{apsrev4-1}

\title{Prediction of two-dimensional ferromagnetic VO$_2$ layers in the hexagonal and tetragonal phases}

\author{Lihui Han}
\affiliation{College of Physics and Engineering, Qufu Normal University, Qufu, Shandong 273165, China}
\author{Lujia Tian}
\affiliation{College of Physics and Engineering, Qufu Normal University, Qufu, Shandong 273165, China}
%\author{Yuanfang Yue}
%\affiliation{College of Physics and Engineering, Qufu Normal University, Qufu, Shandong 273165, China}
\author{Bing-Xin Liu}
\affiliation{School of Physics and Electronics, Shandong Normal University, Jinan 250358, China}
\affiliation{College of Physics and Engineering, Qufu Normal University, Qufu, Shandong 273165, China}
\author{Zong-liang Li}
\affiliation{School of Physics and Electronics, Shandong Normal University, Jinan 250358, China}
\author{Miao Gao}
\affiliation{Department of Physics, School of Physical Science and Technology, Ningbo University, Zhejiang 315211, China}
\author{Fengjie Ma}
\email{fengjie.ma@bnu.edu.cn}
\affiliation{The Center for Advanced Quantum Studies and School of Physics and Astronomy, Beijing Normal University, Beijing 100875, China}
\affiliation{Key Laboratory of Multiscale Spin Physics (Ministry of Education), Beijing Normal University, Beijing 100875, China}
\author{Xun-Wang Yan}
\email{yanxunwang@163.com}
\affiliation{College of Physics and Engineering, Qufu Normal University, Qufu, Shandong 273165, China}

\date{\today}

\begin{abstract}
Ferromagnetism in the two-dimensional materials is of great significance and has become an emerging topic.
The ferromagnetic VS$_2$ and VSe$_2$ monolayers have been experimentally synthesized, and O element belongs to the same group as S and Se elements. Thus, whether there exists the ferromagnetic VO$_2$ monolayer is a necessary and urgent question.
Using first-principles methods within the framework of density functional theory, we predict two kinds of VO$_2$ monolayers with the hexagonal and tetragonal phases and investigate their structural stability, electronic and magnetic properties, and ferromagnetic phase transition. The computational results demonstrate that the two two-dimensional structural phases are stable and possess the ferromagnetic ground states, and they are half-metal with large energy gap. In addition, by solving the Heisenberg model with the Monte Carlo simulation methods, the ferromagnetic phase transition at 270 K in the hexagonal phase is determined.
These findings not only predict a new type of intrinsic half-metallic ferromagnet with a high Curie temperature but also fill in an important gap that are lacking in the series of studies from VO$_2$, VS$_2$, VSe$_2$, to VTe$_2$.

\end{abstract}

\maketitle
\section{Introduction}
Since the discovery of graphene in 2004~\cite{Novoselov2004}, the exploration of novel two-dimensional (2D) materials has become a leading topic in physics and materials. Compared with bulk materials, 2D materials have numerous advantages because of the almost complete exposure of atoms on their surfaces, high atom utilization rate, and ease of regulation~\cite{Zhang2015}. Among 2D materials, transition metal dichalcogenides (TMDs) possess excellent properties in terms of electronic structures, electrocatalytic properties, and optical properties~\cite{Liu2023,Vutukuru2021}, which makes them a highly sought-after research issue. Nevertheless, the existence of magnetism referred to spin degree has not been observed in most of the present 2D TMDs, which seriously hinders their practical applications in the field of information storage, thus the search for 2D TMDs with ferromagnetic orders is of great significance.

Recently, the existence of ferromagnetism in some TMDs monolayers has been demonstrated both in experiments and theories~\cite{Sheng2021,Yang2023,Lu2024,Jimenez2021}.
Firstly, Ma et al. predicted that the VS$_2$ monolayer was a ferromagnetic monolayer through first-principles calculations~\cite{Ma2012}. Based on the work of Ma et al., Gao et al. synthesized ultrathin VS$_2$ nanosheets in experiments~\cite{Gao2013} and verified their ferromagnetism.
Secondly, the VSe$_2$ monolayer was also successfully synthesized~\cite{Yu2019,Bonilla2018}, and Batzill et al. demonstrated that monolayer VSe$_2$ possessed strong ferromagnetism with the Curie temperature higher than room temperature experimentally~\cite{Bonilla2018}.
Thirdly, for the VTe$_2$ monolayer and bilayer, the predicted Curie temperature of ferromagnetic VTe$_2$ monolayer was about 120 K~\cite{Jafari2023}.
O element, being in the same group as S, Se, and Te, has the same number of valence electrons as them. Naturally, we can hypothesize that VO$_2$ monolayer should possess ferromagnetic properties similar to VS$_2$ and VSe$_2$ monolayer. So far, two-dimensional VS$_2$ and VSe$_2$ have been extensively studied~\cite{Pathirage2023,Rajka2022,Wines2023,Yadav2024}, however there are few reports on two-dimensional VO$_2$.

Mermin-Wagner theorem states that there is absence of ferromagnetism and antiferromagnetism in one- or two-dimensional isotropic Heisenberg models, which gives a false impression that long-range magnetic order do not exist in two-dimensional materials and then hampers the research development in the field of magnetic two-dimensional materials~\cite{Mermin1966}. Until 2017, ferromagnetism in the two-dimensional materials was firstly realized in the CrI$_3$ layers with single-ion magnetic anisotropy in experiments~\cite{Huang2017}. Since then, two-dimensional magnetic materials are intensively investigated by the theoretical and experimental researchers~\cite{Samarth2017,Burch2018,Shabbir2018} and our group also predicted a few magnetic transition metal nitride or carbonitride monolayers~\cite{Liu2021a,Liu2021b,Zhang2022,Liu2022}. However, up to now, the variety of magnetic two-dimensional materials synthesized experimentally is still very few. Therefore, the exploration of new magnetic two-dimensional materials, especially in the metal oxides, is an important and necessary research topic.
On the other hand, bulk transition metal oxides are usually strongly correlated electron systems and can exhibit fascinating physical properties, including magnetoresistance, superconductivity, and metal-insulator transition~\cite{Ahn2021,Lee2020}.
There are the strong electron-lattice interactions and electron-electron correlations in the bulk VO$_2$ compound. It is a well-established chromogenic material and can respond to electromagnetic radiation, temperature, and electrical charge~\cite{Bahlawane2014}. The insulator-to-metal transition and its applications have also  been intensively investigated in the previous studies~\cite{Morin1959}.
When bulk transition metal oxides are reduced to two-dimensional compounds, it is doubtful whether their structure will remain stable and whether physical properties will be maintained. Hence, two-dimensional VO$_2$ is worth studying.

In this work, based on the first-principles methods in the framework of density-functional theory, we investigate the hexagonal and tetragonal phases of VO$_2$ monolayers, and the results indicate that these two phases possess dynamical, thermal, and mechanical stability. Spin polarization calculations demonstrate that the hexagonal and tetragonal phases are 2D ferromagnetic half-metals. The Curie temperature is determined by solving the Heisenberg model with Monte Carlo simulations, and the results demonstrate that the hexagonal phase has a higher Curie temperature.

%\captionsetup[figure]{name={Fig.},labelsep=period}
\begin{figure}[H]
\centering
\includegraphics[width=8.0cm]{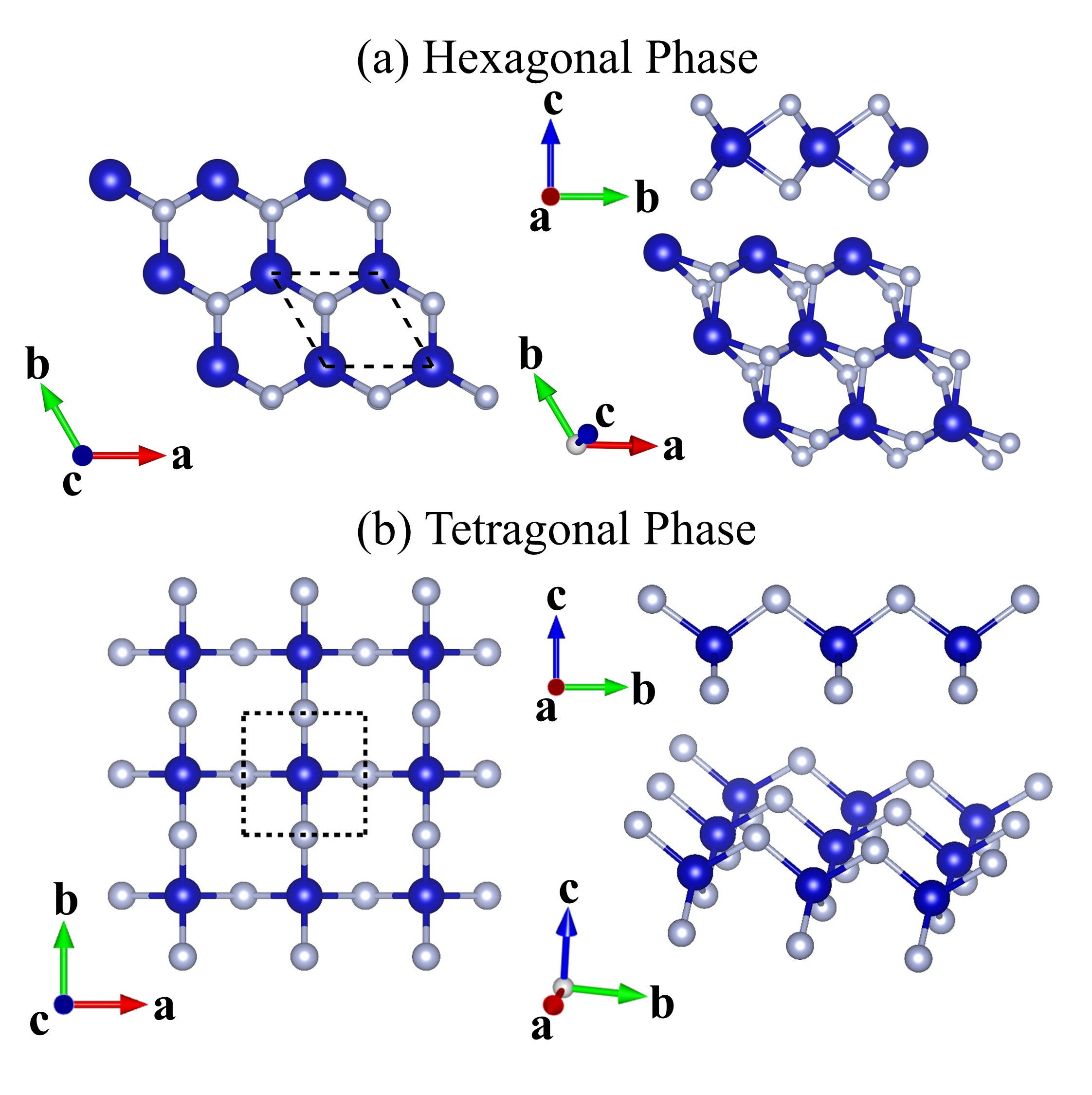}
\caption{Top view, side view, and oblique view of the VO$_2$ structures in the hexagonal (a), and tetragonal (b) phases. The unit cell is marked with a black dashed line. The dark blue ball represents the V atom and the silver ball represents the O atom.}
\label{fig.1}
\end{figure}

\section{COMPUTATIONAL DETAILS}

The calculations are performed in the VASP package (Vienna ab-inito simulation package)~\cite{PhysRevB.47.558,PhysRevB.54.11169}. The projector augmented-wave method (PAW)~\cite{PhysRevB.50.17953} as well as the generalized gradient approximation (GGA) with Perdew-Burke-Ernzerhof (PBE) formula~\cite{PhysRevLett.77.3865} for ionic potential are employed. The GGA + \textit{U} method is applied to consider the correction of electron correlation~\cite{Dudarev1998}. The plane wave basis cutoff is 600 eV and the convergence thresholds for the total energy and force were 10$^{-5}$ eV and 0.01 eV/\AA. The interlayer distance is set to 16 \AA, and a mesh of 28 $\times$ 28 $\times$ 1 k-points is used for the Brillouin zone integration. The phonon calculations are carried out with the supercell method in the PHONOPY program, and the real-space force constants of supercells are calculated using density functional perturbation theory (DFPT) as implemented in VASP~\cite{Togo2015}. To inspect the thermal stability of hexagonal and tetragonal phases of VO$_2$, the ab initio molecular dynamics (AIMD) simulations are performed. The 5 $\times$ 5 $\times$ 1 supercells are employed to minimize the constraint of periodic boundary conditions and the temperature is kept at 1000 K for 5 ps with a time step of 1 fs in the moles-volume-temperature (NVT) ensemble~\cite{Martyna1992}. The temperature of the phase transition in the hexagonal and tetragonal phases of VO$_2$ is evaluated by solving the Heisenberg model, and the 100 $\times$ 100 lattice is used in the Monte Carlo simulation.

\section{Results and Discussions}
\subsection{Atomic structure}

The atomic structures of the VO$_2$ monolayer in the hexagonal and tetragonal phases are shown in Fig.\ref{fig.1}. The unit cell is marked with a black dashed line. The unit cell of VO$_2$ is composed of one V atom and two O atoms, represented by dark blue and silver ball, respectively. The hexagonal phase has the symmetry of P$\bar{6}$m2 (No. 187) space group, in which the V atom is coordinated with six O atoms, located at the center of a regular triangular prism formed by six O atoms. The V-O distances is 1.95 $\AA$ and the O-O distances is 2.27 $\AA$. The symmetry of tetragonal phase belongs to the P$\bar{4}$m2 space group (No. 115). In the tetragonal phase, the four O atoms form a tetrahedron and V atom sit at the center of tetrahedron. The VO$_4$ tetrahedrons are linked together to make up a square network by sharing O atoms. The V-O distances is 1.80 \AA.
Usually, two-dimensional transition metal dichalcogenides have multiple structural phases, including trigonal phase, hexagonal phase, and tetragonal phase. For the VO$_2$ monolayer, only hexagonal and tetragonal phases are considered in this work because the trigonal phase is demonstrated to be unstable in our phonon spectra calculations.

\subsection{Structural stability}

%\captionsetup[figure]{name={Fig.},labelsep=period}
\begin{figure}[H]
\centering
\includegraphics[width=8.5cm]{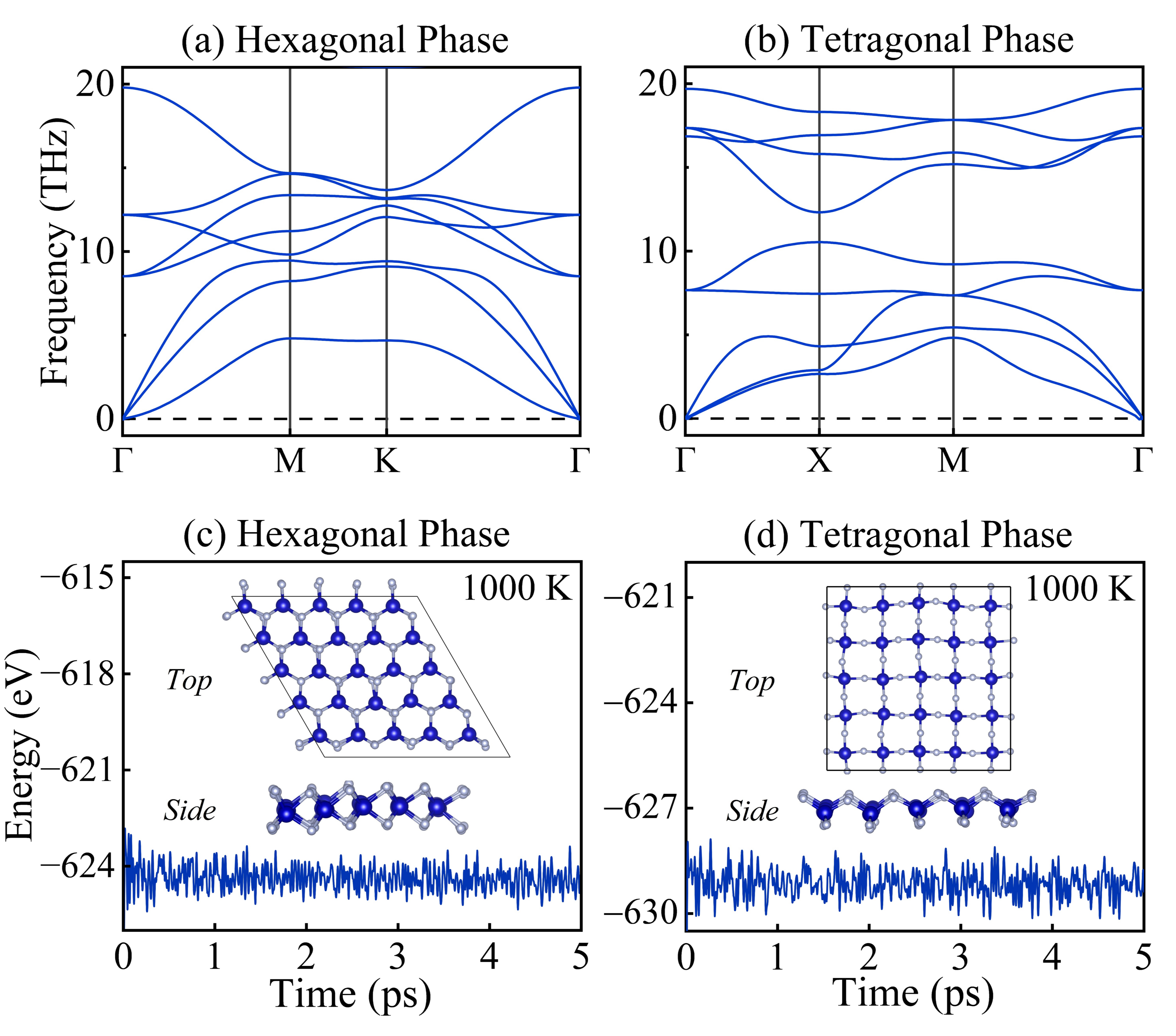}
\caption{(a) and (b), Phonon spectra of hexagonal phase and tetragonal phase. (c) and (d), Evolution of total potential energy with the time at 1000 K for 5 ps, where the insets are the top and side views of the final configuration after molecular dynamics simulations.}
\label{fig.2}
\end{figure}

We perform calculations of the phonon spectra, molecular dynamics simulations, elastic constants, and formation energy to verify the structural stability of VO$_2$ in the hexagonal and tetragonal phases. Fig. \ref{fig.2}(a) and (b) display the phonon spectra for the hexagonal and tetragonal phases, respectively. For the hexagonal phase, the high-symmetry points are $\Gamma$(0 0 0), M(0.5 0 0), and K(1/3 1/3 0) in reciprocal space, and for the tetragonal phase, the high-symmetry points are $\Gamma$(0 0 0), X(0.5 0 0), and M(0.5 0.5 0). There is no imaginary frequency in the phonon spectra, indicating that hexagonal phase and tetragonal phase structure are dynamically stable. After that, we adopt the molecular dynamic simulations at 1000 K for 5 ps to examine the thermal stability of hexagonal phase and tetragonal phase, which are shown in Fig. \ref{fig.2}(c) and (d). It can be seen that the total potential energy fluctuates around a certain value and there is no significant drop in energy. The insert is the snapshot of final structure after the molecular dynamic simulation, which displays that the framework is maintained without broken bonds. The results from the molecular dynamic simulations indicate that hexagonal phase and tetragonal phase possess good thermal stability.

In order to further investigate the mechanical stability of VO$_2$ structures in the hexagonal and tetragonal phases, we compute the elastic constants and obtain the values, as shown in Table \ref{Table.1}. These elastic constants satisfy the two inequalities $C_{11}C_{22}-C_{12}C_{21} > 0$ and $C_{66} > 0$ , namely, in agreement with the mechanical stability Born criteria~\cite{Born:224197}. Hence, the hexagonal and tetragonal phases have good mechanical stability.
\begin{table}[] \centering
\caption{Elastic constants of VO$_2$ structures in the hexagonal and tetragonal phases. The unit is N/m.}
\renewcommand\tabcolsep{6.5pt}
\renewcommand\arraystretch{1.5}
\begin{tabular*}{8.5cm}{cccccc} \hline\hline
      & ${C}_{11}$   & ${C}_{22}$   & ${C}_{12}$ & ${C}_{12}$  & ${C}_{66}$   \\ \hline
Hexagonal & 200.67 & 200.67 & 76.64 & 76.64 & 63.55  \\
Tetragonal & 84.58 & 84.58 & 5.06 & 5.06 & 8.82 \\ \hline\hline
\end{tabular*}
\label{Table.1}
\end{table}

\begin{figure}[H]
\centering
\includegraphics[width=8.5cm]{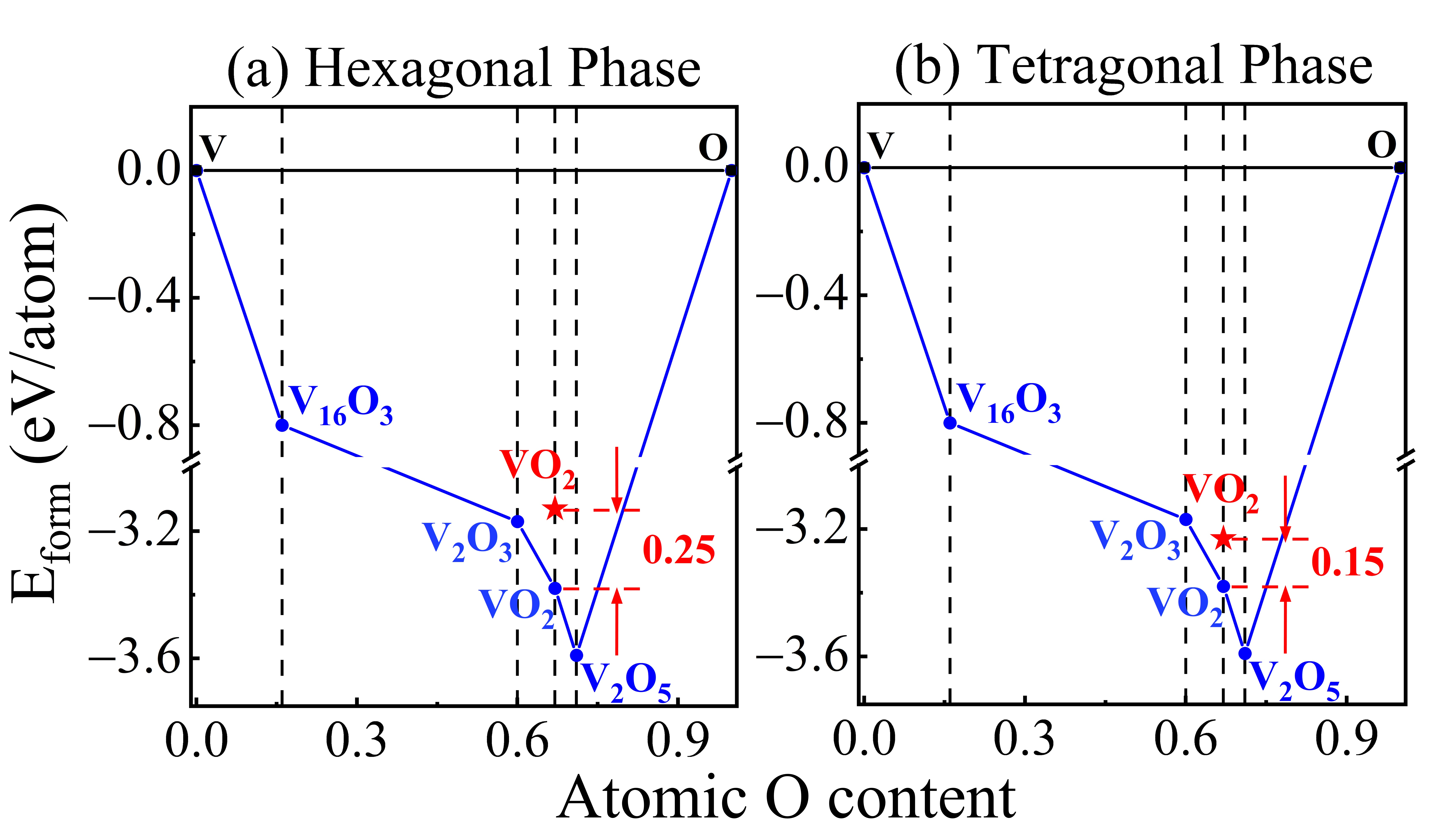}
\caption{The convex hull of formation energy of VO$_2$ structures in the hexagonal phase (a) and tetragonal phase (b).}
\label{fig.3}
\end{figure}

Moreover, the formation energies of the hexagonal and tetragonal phases are calculated according to \begin{equation}
    {E}_{form}=({E}_{tot}-{E}_{metal}-2{E}_{{O}})/3
\end{equation}
in which \textit{E}$_{tot}$, \textit{E}$_{metal}$ and \textit{E}$_{O}$ are the total energy, bulk metal energy per atom, and elemental O energy. The formation energy averaged over atoms in hexagonal and tetragonal phases are -3.13 eV/atom and -3.23 eV/atom, respectively.
For comparison, we plot the binary phase diagrams of V-O compounds. As shown in Fig. \ref{fig.3}, bulk V$_{16}$O$_3$~\cite{Hiraga1973}, V$_2$O$_3$~\cite{Rozier2002}, VO$_2$~\cite{Andersson1956}, V$_2$O$_5$~\cite{Shklover1996}, metal V~\cite{Karen2005}, and elemental O make up the V-O convex hull. The hull energies of hexagonal and tetragonal phases are only 0.25 eV/atom and 0.15 eV/atom, respectively. Due to different compositional ratio and structural isomer, there are a variety of binary vanadium oxides. In Ref.~\citenum{Bahlawane2014}, the reported energy above the convex hull line is 0.1857 eV/atom for VO$_2$ in the trigonal phase with R3m group symmetry and the hull energy of V$_7$O$_3$ in monoclinic (C2/m) structure reaches up to 0.800 eV/atom.
Therefore, the hull energies for two VO$_2$ monolayers predicted are relatively small, indicating that it is feasible to fabricate the hexagonal and tetragonal VO$_2$ monolayers in experiments.

%\captionsetup[figure]{name={Fig.},labelsep=period}

\subsection{Electronic structure}

In the 3$d$ transition metal compounds, the electron-electron correlation is difficult to be described due to the complex $d$ electrons interaction. The GGA + \textit{U} method provides a more precise description of the electronic correlation than the pure PBE functionals method, in which the Hubbard \textit{U} parameter is introduced to account for the correlation of the onsite Coulomb interaction. In our calculations, the GGA + \textit{U} method with the Hubbard \textit{U} value of 3.10 eV is adopted~\cite{Jain2011}.

\begin{figure}[H]
\centering
\includegraphics[width=8.5cm]{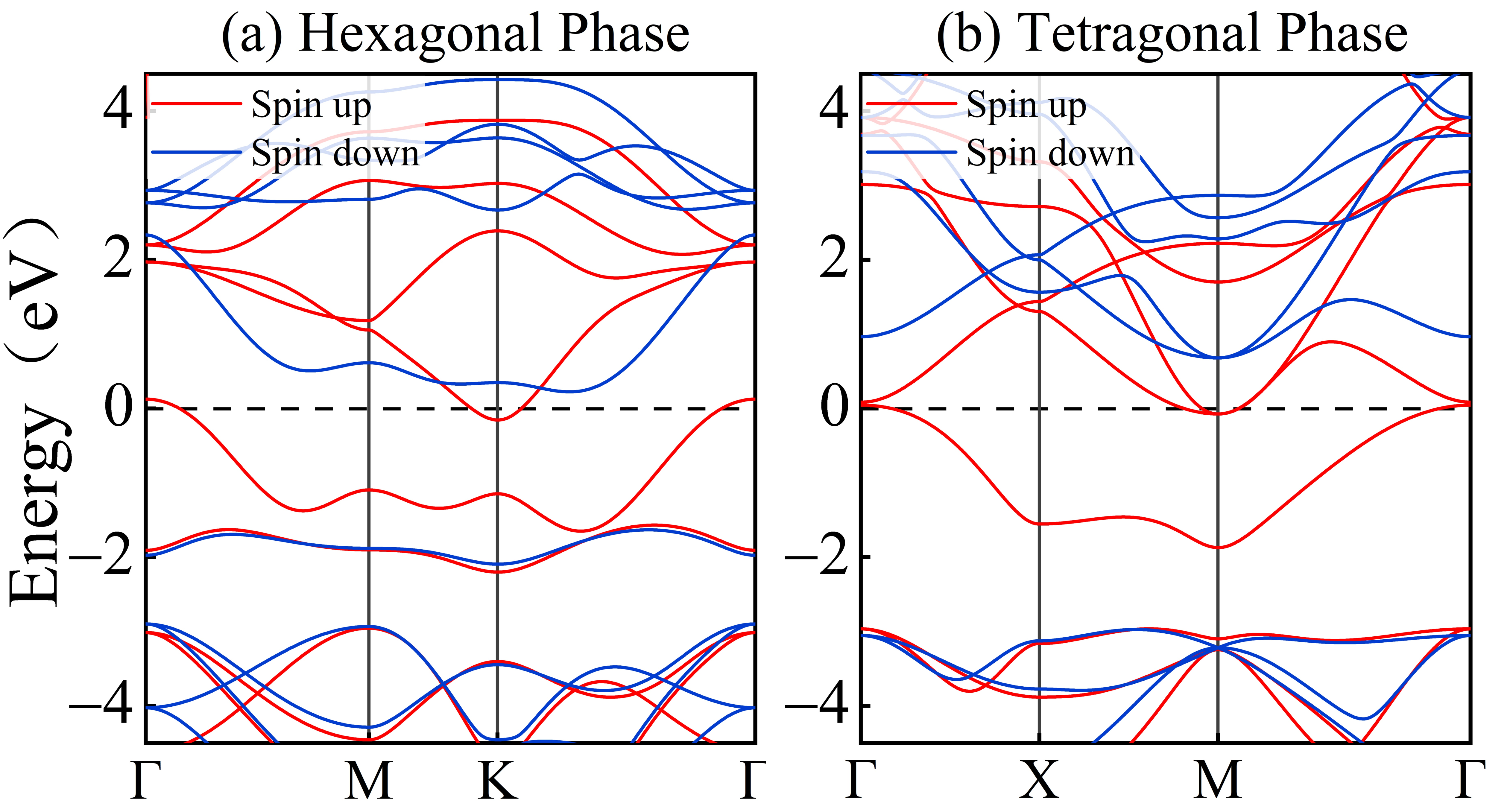}
\caption{Band structure of the VO$_2$ structures in the hexagonal and tetragonal phase from the GGA + \textit{U} calculations. Fermi energy is set to zero. (a) hexagonal phase, (b) tetragonal phase.}
\label{fig.4}
\end{figure}
Fig.\ref{fig.4} displays the energy bands of the hexagonal and tetragonal VO$_2$ monolayers in the ferromagnetic (FM) order. The red and blue curves correspond to the spin-up and spin-down bands. For the hexagonal phase, the highly symmetry points are $\Gamma$(0 0 0), M(0.5 0 0), and K(1/3 1/3 0) in reciprocal space, and for the tetragonal phase, the high-symmetry points are $\Gamma$(0 0 0), X(0.5 0 0), and M(0.5 0.5 0).
In Fig.\ref{fig.4} (a) and (b), the spin-up and spin-down bands appear an obvious spin splitting. Among them, the spin-up bands cross the Fermi level and take on the metallic features, while the spin-down bands form a energy gap near the Fermi energy, exhibiting the properties of semiconductors or insulators~\cite{Graf2011}. The hexagonal and tetragonal phases of VO$_2$ monolayers are fine half-metal with the gaps as high as 1.86 and 3.65 eV, which can effectively prevent the spin-flip caused by thermal perturbation~\cite{de1983}.

\begin{figure}[H]
\centering
\includegraphics[width=7.5cm]{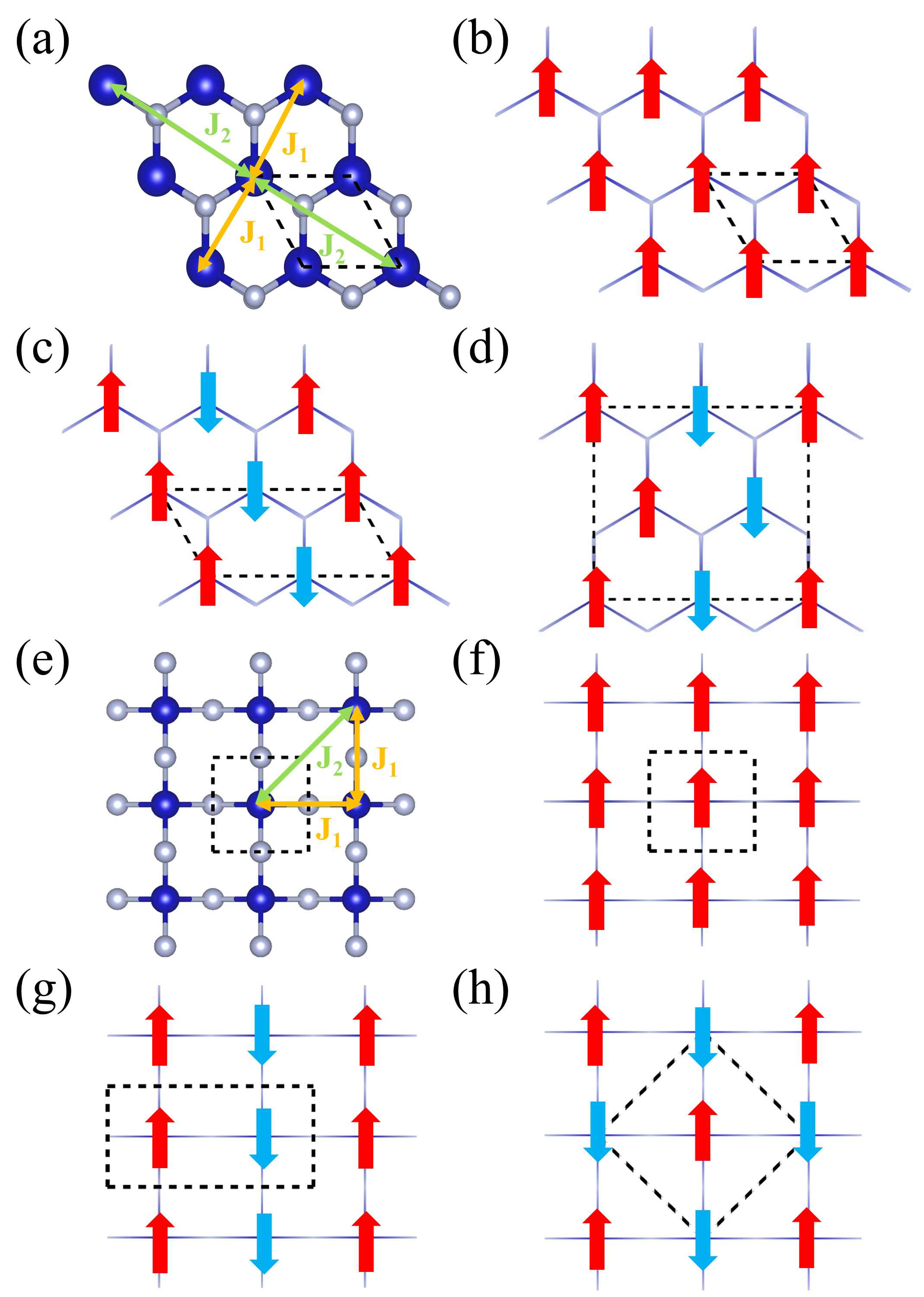}
\caption{Hexagonal phase: (a) Atomic structure,  (b) FM order, (c) AFM-I order (d) (AFM-II) order. Tetragonal phase: (e) Atomic structure, (f) FM order, (g) AFM-I order (h) (AFM-II) order. $J_1$ and $J_2$ are the nearest and next-nearest neighbor exchange couplings, represented by the double arrows.  The magnetic unit cells are marked with the dashed lines. Red and blue arrows indicate the opposite directions of V magnetic moments. For convenience to observe the magnetic order, the atomic structure is displayed with the wire frame.}
\label{fig.5}
\end{figure}

\subsection{Ferromagnetic ground state}
Next, we inspect the magnetic ground states of VO$_2$ structures in the hexagonal and tetragonal phases.
For the two structural phases, Fig. \ref{fig.5} displays the top view of structure and three kinds of magnetic order including the ferromagnetic order (FM), antiferromagnetic order I (AFM-I), and antiferromagnetic order II (AFM-II).
In them, the magnetic unit cells are marked with the dashed lines, and red and blue arrows indicate the opposite directions of magnetic moments around V atoms. $J_1$ and $J_2$ are the nearest and next-nearest neighbor exchange couplings.
We compute the energies of different magnetic orders and list the values in Table \ref{TABLE.2}.
For clarity, the FM energy is set to 0 meV, and the energies of AFM-I and AFM-II are the ones relative to the FM energy.
The results show that the FM order has the lowest energy in both the hexagonal and tetragonal phases, demonstrating that the two transition metal oxide monolayers are ferromagnetic two-dimensional materials.
 %, in which a single V atom has magnetic moment of 1.07 $\mu_B$.
\begin{table*}\centering
\caption{The energies of hexagonal phase and tetragonal phase in the FM, AFM-I, and AFM-II orders, the nearest and next-nearest neighbor exchange couplings $J_1$ and $J_2$, the magnetic anisotropic energy $A$ , the Curie temperature T$_C$. The FM energy is set to zero and the energy is the one per formula cell.}
\renewcommand\tabcolsep{5pt}
\renewcommand\arraystretch{1.5}
\begin{tabular*}{17.8cm}{ccccccccc} \hline\hline
      & $ E_{FM}$   (meV) & E$_{AFM-I}$   (meV) & E$_{AFM-II}$   (meV) & Moment   ($\mu_B$) & J$_1$ (meV/S$^2$) & J$_2$ (meV/S$^2$) & $A$ (meV/S$^2$) & T$_C$ (K) \\ \hline
Hexagonal phase &  0     & 87.06    & 87.12      & 1.05     & -21.79    & 0.03     & -0.67    & 270 \\
Tetragonal phase &  0     & 9.15    & 13.57     & 1.06     & -3.39    & -0.59     & -0.004    & 27
\\ \hline\hline
\end{tabular*}
\label{TABLE.2}
\end{table*}

\subsection{Curie temperature}

The Curie temperature is a key parameter to determine the practical application of the magnetic materials. We employ the two-dimensional Heisenberg model to describe the magnetic interactions in the hexagonal and tetragonal phases. The Hamiltonian is defined as
\begin{equation}
    H={J}_{1}\displaystyle\sum_{<ij>}\overrightarrow{S}_{i}\cdot\overrightarrow{S}_{j} +{J}_{2}\displaystyle\sum_{<<ij^{\prime}>>}\overrightarrow{S}_{i}\cdot\overrightarrow{S}_{j^{\prime}}+A\displaystyle\sum_{i}(S_{iz})^2
\end{equation}
in which j and j$^{\prime}$ denote the nearest and next-nearest neighbors of i site. $J_1$ and $J_2$ are the nearest and next-nearest neighbor exchange couplings, respectively. $A$ is the magnetic anisotropic energy, and the relevant data are shown in Table \ref{TABLE.2}.
The exchange couplings $J_1$ and $J_2$ are derived from the difference of energies shown in TABLE \ref{TABLE.2}, according the Formula \ref{Eq1} and \ref{Eq2} below~\cite{Ma2008},
\begin{equation} \label{Eq1}
  \begin{aligned}
{J}_{1}&=({E}_{FM}+{E}_{AFM-I}-2{E}_{AFM-II})/4,  \\
{J}_{2}&=({E}_{AFM-II}-{E}_{AFM-I})/2
  \end{aligned}
\end{equation}

\begin{equation} \label{Eq2}
  \begin{aligned}
{J}_{1}&=({E}_{FM}-{E}_{AFM-II})/4,  \\
{J}_{2}&=({E}_{FM}-2{E}_{AFM-I}+{E}_{AFM-II})/8
  \end{aligned}
\end{equation}

\begin{figure}[H]
\centering
\includegraphics[width=8.5cm]{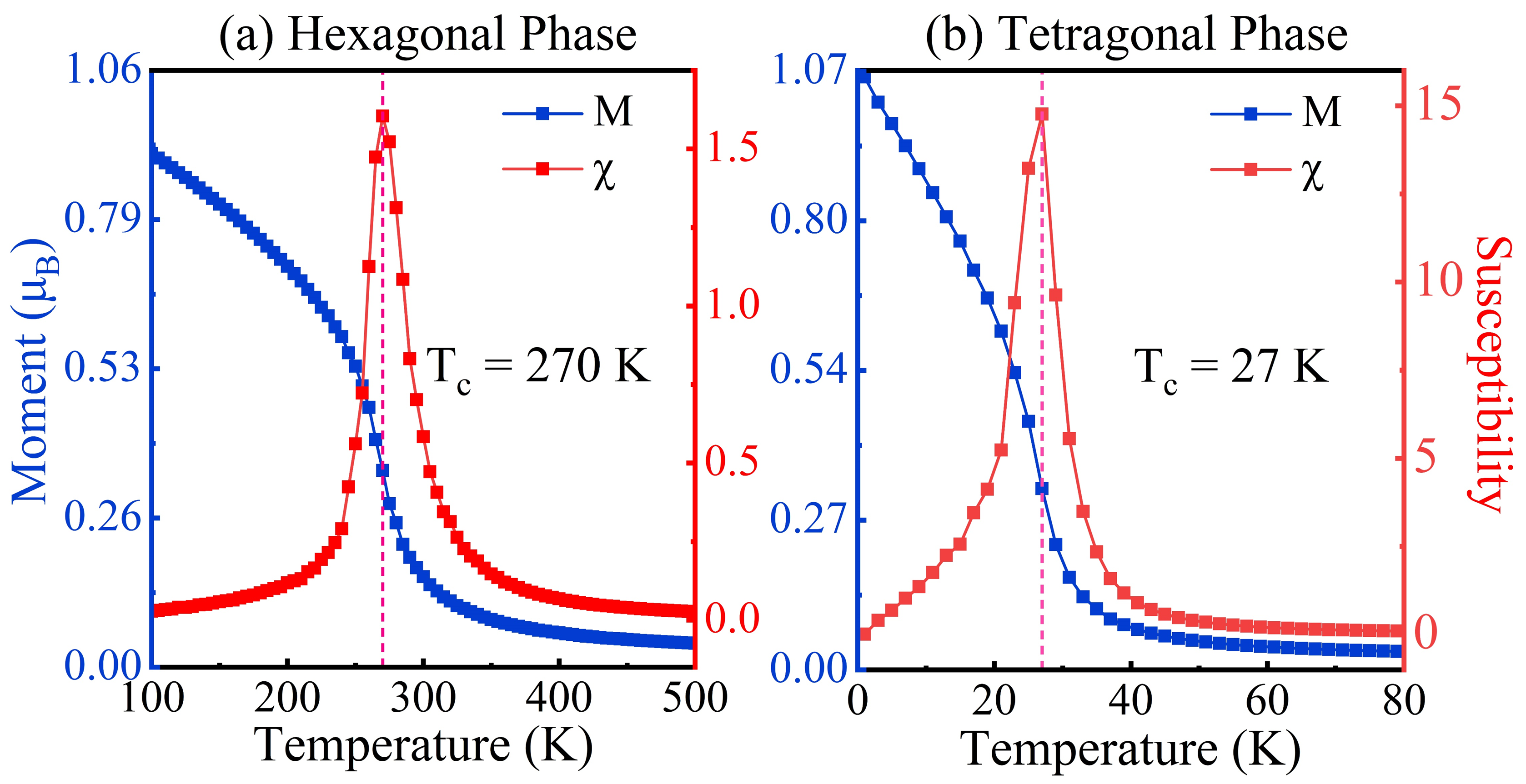}
\caption{The variation of average moment and susceptibility ($\chi$) with temperature. (a) hexagonal phase, (b) tetragonal phase.}
\label{fig.6}
\end{figure}

Then, to determine the Curie temperature of the hexagonal and tetragonal phase, we solve the Hamiltonian of Heisenberg model by Monte Carlo method. The variation of average magnetic moment (\textit{M}) and susceptibility with temperature are presented in Fig.\ref{fig.6}. The Curie temperature is 270 K for hexagonal phase and the Curie temperature is 27 K for tetragonal phase. The result shows that the VO$_2$ monolayer in the hexagonal phase is not only an intrinsic layered half-metal with a wide band gap, but also has high Curie temperature, which endows it potential applications in spintronics.

\section{Conclusions}

Based on the first-principles approach within the framework of density functional theory, we have predicted VO$_2$ structures in the hexagonal and tetragonal phases. Phonon spectral calculations, molecular dynamics simulations, and elastic constant calculations have been conducted to determine that hexagonal and tetragonal phases are dynamically, thermally and mechanically stable. Moreover, formation energy and convex hull calculations further indicate the high possibility of synthesizing these two structures in experiments. Band calculations indicate that two structures are 2D magnetic half-metals with a wide band gap. By comparing the energies at different magnetic orders, the ferromagnetism is determined in the ground state of VO$_2$ structures in the hexagonal and tetragonal phases. Based on the Heisenberg model with Monte Carlo simulations, we find that the hexagonal phase has a high Curie temperature.
These findings not only predict a new type of intrinsic half-metallic ferromagnet with a high Curie temperature but also provide an important complement for the series studies on the two-dimensional materials VS$_2$, VSe$_2$, and VTe$_2$.

\begin{acknowledgements}
This research was funded by the National Natural Science Foundation of China under Grants Nos. 12274255, 12074040, 11974194, 12474286, 11974207 and the Major Basic Program of Natural Science Foundation of Shandong Province under Grant No. ZR2021ZD01. M. Gao was also supported by Ningbo Youth Science and Technology Innovation Leading Talent Project (No. 2023QL016) and F. Ma was also supported by the BNU Tang Scholar.
\end{acknowledgements}

\bibliography{library}%
\iffalse

\fi
\end{document}